\documentclass[a4paper]{article}
\usepackage{amsmath}
\usepackage{amsfonts}
\usepackage{url}
\usepackage{verbatim}
\usepackage{algorithm} 
\usepackage{algorithmic} 
\usepackage[dvipdfm]{graphicx}        
\usepackage{abstract}

\newtheorem{theorem}{Theorem}[section]

\newtheorem{definition}[theorem]{Definition}

\def\F2{{\mathbb F}_2}
\def\ba{{{\mathbf a}}}

\def\bx{{{\mathbf x}}}
\def\by{{{\mathbf y}}}

\def\bt{{{\mathbf t}}}
\def\bu{{{\mathbf u}}}
\def\bo{{{\mathbf o}}}
\def\b1{{{\mathbf 1}}}
\def\bA{{{\mathbf A}}}

\def\bT{{{\mathbf T}}}

\def\bR{{{\mathbf R}}}
\def\bzero{{{\mathbf 0}}}

\title{Variants of Mersenne Twister Suitable for Graphic Processors}
\author{MUTSUO SAITO\\Hiroshima University \and
MAKOTO MATSUMOTO\\The University of Tokyo}

\begin{document}
\maketitle

\begin{abstract}
This paper proposes a type of pseudorandom number generator,
{\em Mersenne Twister for Graphic Processor} (MTGP),
for efficient generation on graphic processessing units (GPUs).
MTGP supports large state sizes such as 11213 bits, and
uses the high parallelism of GPUs in computing many steps of
the recursion in parallel.
The second proposal is a parameter-set generator for MTGP,
named {\em MTGP Dynamic Creator} (MTGPDC).
MTGPDC creates up to $2^{32}$
distinct parameter sets which generate sequences with high-dimensional uniformity.
This facility is suitable for
a large grid of GPUs where each GPU requires separate random number streams.

MTGP is based on linear recursion over the two-element field,
and has better high-dimensional equidistribution
than the Mersenne Twister pseudorandom number generator.
\end{abstract}

keywords: Pseudo Random Number Generator, Mersenne Twister,
General-Purpose Computing on Graphics Processing Units,
Dynamic Creator







%
%
\section{Random number generation for GPU}
\label{sec:introduction}
A Graphic Processing Unit (GPU) is a highly parallel processor designed for
computer graphics. GPUs are now widely used in both personal computers
and game machines, and are cheap despite their high computational power.
As a consequence, a trend in parallel computation is
{\em General-Purpose computing on Graphics Processing Units (GPGPU)}~\cite{GPGPU},
namely, to utilize the high parallelism of GPUs for
solving computing problems outside the graphics domain.
A number of recent super-computers actually consist of a large grid
of GPUs, controlled by one or several CPUs.

Pseudorandom number generators are often necessary in GPGPU,
for example, in Monte Carlo simulations,
so it is useful to design pseudorandom number generators
taking advantage of the parallelism of GPUs.

We propose a class of pseudorandom number generators,
{\em  Mersenne Twister
for Graphic Processors} (MTGP).
The algorithm is similar to that of Mersenne Twister (MT) \cite{MT},
but refined and adjusted to the hierarchical parallelism of GPUs.
The parameter sets for generators are selected by their
high-dimensional equidistribution properties.
We prepared 128 different parameter sets for each
of the periods $2^{11213}-1$,
$2^{23209}-1$ and $2^{44497}-1$.
The gap between the theoretical upper bound and the realized dimensions
of equidistribution of MTGPs is smaller than those of MT,
so the MTGPs provide better
statistical quality for the same amount of storage.

We also designed a parameter-set generator for MTGP,
named MTGP Dynamic Creator (MTGPDC). Analogously to
Dynamic Creator \cite{DC} for MT, MTGPDC finds
good parameter-sets for MTGP
with high dimensions of equidistribution.
Users specify a set of generator IDs and the
desired Mersenne prime period, then the ID is
embedded in the recursion parameters of each
generator,
so that generators with distinct IDs will yield independent streams.
This is useful for large grids of GPUs, where each
GPU needs one or more independent random
number streams, as each stream can be generated
from generator recursion parameters with a distinct ID.

The design policies of MTGP and MTGPDC are as follows.
\begin{enumerate}
\item
Many parallel threads operate on
the state space of one large pseudorandom number generator (PRNG).
The large state allows a long period
and high-dimensional equidistribution properties.
This is in contrast to the usual approach to PRNG parallelism, namely,
one generator per thread. In the latter case,
the increase in the number of threads implies
that each generator has a small state space, since
the size of fast memory in a GPU is limited.

\item
GPUs have a hierarchy in memory: some memory is fast but
its access is limited to a group of threads (called a {\em block}),
and some memory is fast but read-only. MTGP takes into account the
characteristics of
each class of memory for efficient parallel generation.
\end{enumerate}

%
%
\section{GPGPU and CUDA}
\label{sec:cuda}

For GPGPU, a typical hardware setting is as follows.
One CPU, called {\em the host CPU}, is
connected to one GPU (often to a grid of GPUs, but for simplicity
of explanation we choose the one-GPU case). Consider a computation $C$ which
one wants to do in GPGPU. As usual, $C$ is divided into several parts,
and some of the parts can be executed in parallel.
To use the GPU effectively, one needs to analyze which part of the
program can run in parallel, in the many processing units in the GPU.
Then, one writes a program for the CPU, called {\em the host program},
which sends input data and GPU codes (called {\em the kernel program})
to the GPU for parallel execution.
The GPU does the given computation, and returns the output data to the CPU.
Usually the output data of GPU computation is a large array, so
the data is written in a memory called {\em global memory}
which is accessible from the CPU.

Under this setting, the computer program must be equipped with facilities
for scheduling execution of GPU kernels and
communication between the CPU and the GPU.
One solution is the
{\em Compute Unified Device Architecture} (CUDA)~\cite{CUDA1},\cite{CUDA2},
which is a widely used software-development environment for GPGPU
on NVIDIA's GPUs.
CUDA has a C-like language which supports these facilities.
A CUDA program consists of one host program
and several kernel programs.
Some more concrete explanation is in \S~\ref{app:C}.

MTGP is written in
and executed in the CUDA-environment.
Initialization of MTGP is done in the host program (i.e. by the CPU),
and the host program calls a kernel program (i.e.\ invokes the GPU)
which generates a specified number of pseudorandom numbers in
the global memory (by the GPU).
MTGPDC is written in C and executed in the usual CPU (i.e. without GPU).

Another programming environment for GPGPU is the OpenCL \cite{OPENCL}.
At present, there is no OpenCL version of MTGP/MTGPDC. We use CUDA's terminology,
but put comments on OpenCL's corresponding terminology when they differ.

\subsection*{Hierarchical Structure: software and memory}
Since the hardware architecture of a GPU is
rather complicated, we explain mainly its software-level
hierarchy.
When we mention concrete values as examples, we use a middle-range GPU
GTX260.

In a kernel program, (namely, a program executed in the GPU),
the minimum execution unit is called a {\em thread} ({\em work item} in OpenCL),
which is one lane
in a CUDA execution grid.
A {\em block} ({\em work group} in OpenCL) consists of many threads,
with some upper bound on their number (e.g. 512 under the present CUDA).
We may think of these threads in one block as running in parallel.
A GPU can run several blocks in parallel.

To realize the high parallelism, threads and blocks have
strong constraints in accessing memory and getting instructions.
There is a corresponding hierarchy in memory.
Each thread has its own set of {\em registers},
which is inaccessible from other threads.
A block has its own {\em shared memory} ({\em local memory} in OpenCL)
of size 16Kbyte (for most CUDA-enabled GPUs at present),
inside the GPU chip. Any thread in one block
can access the shared memory of the block, but can not
access those of other blocks.
A block has also its own read-only cache of the
{\em texture memory} ({\em image object} in OpenCL),
intended for storing texture information in computer graphics.

The tightest restriction is that
any thread in a block gets the same instruction sequence.
Each thread has its own thread-ID number (consecutive)
and can refer to that.
Consequently, each thread acts differently according to its ID-number.

Global memory is stored in external memory chips, located outside the
GPU chip.
Although our ideas on MTGP apply more generally,
for the readers who may want to grasp the size
and speed of GPU,
we give illustrative examples: the size of the memory, in the case of
GTX260, is typically 896Mbyte. Data-transfer speed between global
memory and the GPU
is 112Gbyte/sec, which is faster than typical CPU-memory
transfer speeds (eg.\ 26Gbyte/sec.) Still, the global memory
is slower to access than the shared memory inside the GPU.
Unlike the shared memory,
all threads in the GPU, regardless of which block they belong to, can
access the global memory.
Different blocks can exchange information only via the global memory.
The shared memory is grouped into 16 banks, according
to the address (as memory of 32-bit words) modulo 16.
The threads in one block are grouped in {\em warps}.
A warp consists of 32 threads in the present CUDA-enabled
GPUs.
A half warp (16 threads in a warp) may simultaneously
access the shared memory, only if each thread accesses to
a mutually distinct bank.
If two or more threads in a half warp access
the same bank (namely, the memory addresses coincide modulo 16),
then they can not access in parallel. This phenomenon is called
a {\em bank conflict}.

The cache of the texture memory
can be read by many threads simultaneously,
differently from the shared memory where bank-conflicts may occur.

%
%
\section{Mersenne Twister for GP (MTGP)}
%
%
\subsection{Pseudorandom number generators by recursion}
\label{sec:prng-mtgp}
Let $W$ be the set of $w$-bit ($w$: word size) integers,
and $x_i \in W$ ($i=0,1,2,\ldots$) be a sequence of
$w$-bit integers.

For generating pseudorandom numbers, it is common
to use a recursion:
Let $N$ be a positive integer (called the {\em degree}),
$f:W^N \to W$ be a function, and
$x_0,x_1,\ldots,x_{N-1}$ be elements of $W$ (called the initial values).
Then, the following recursion generates a sequence of elements
in $W$:
\begin{equation}
x_{N+i}:=f(x_{N-1+i},x_{N-2+i}, \ldots, x_i) \ \ (i=0,1,\ldots).
\end{equation}
For high-speed generation, it is better to choose an $f$
depending only on few variables, but if the variables are too few,
the generated sequence tends to show non-randomness.
As a trade-off, we consider the following type of recursion:
\begin{equation}\label{eq:recursion}
x_{N+i}:=f(x_{M+i},x_{1+i},x_i) \ \ (i=0,1,\ldots).
\end{equation}
We call the positive integer $M$ $(1<M<N)$
the {\em middle position}.

Such a recursion can be efficiently computable
(under an appropriate choice of $f$)
using an array $X[0..L-1]$ of words of length $L$ with $L \geq N$
(see \cite[Algorithm A, p.28]{knuth:bible}), as follows.
\begin{enumerate}
\item
Store the initial values $x_0, \ldots, x_{N-1}$
to $X[0], X[1], \ldots, X[N-1]$.

\item
Set an integer variable $i$ to $0$.

\item
\label{item:repeat}
Set
$$X[(N+i) \bmod L] \leftarrow f(X[(M+i) \bmod L],X[(1+i)\bmod L],X[i \bmod L]).$$
This computes $x_{N+i}$.

\item
Increment $i \leftarrow i+1$.
If $i\geq L$, then $i \leftarrow i \bmod L$.

\item
Return to step \ref{item:repeat}.
\end{enumerate}

\subsection{Parallel computation of a single recursion}
Suppose that $L$ is sufficiently large,
namely $L \geq 2N-M$.
Then, one may compute
$$
  X[N+i]\leftarrow f(X[M+i],X[1+i],X[i])
$$
for $0\leq i \leq N-M-1$ in parallel.
When $i=N-M$, the computation of $X[N+i]$ requires
the value of $X[M+i]=X[N]$, which can be obtained
only after the $i=0$-th step has been done, giving
the above upper bound $N-M$ for the number of
parallel threads.

\begin{figure}
  \begin{center}
    \includegraphics[width=0.8\textwidth]{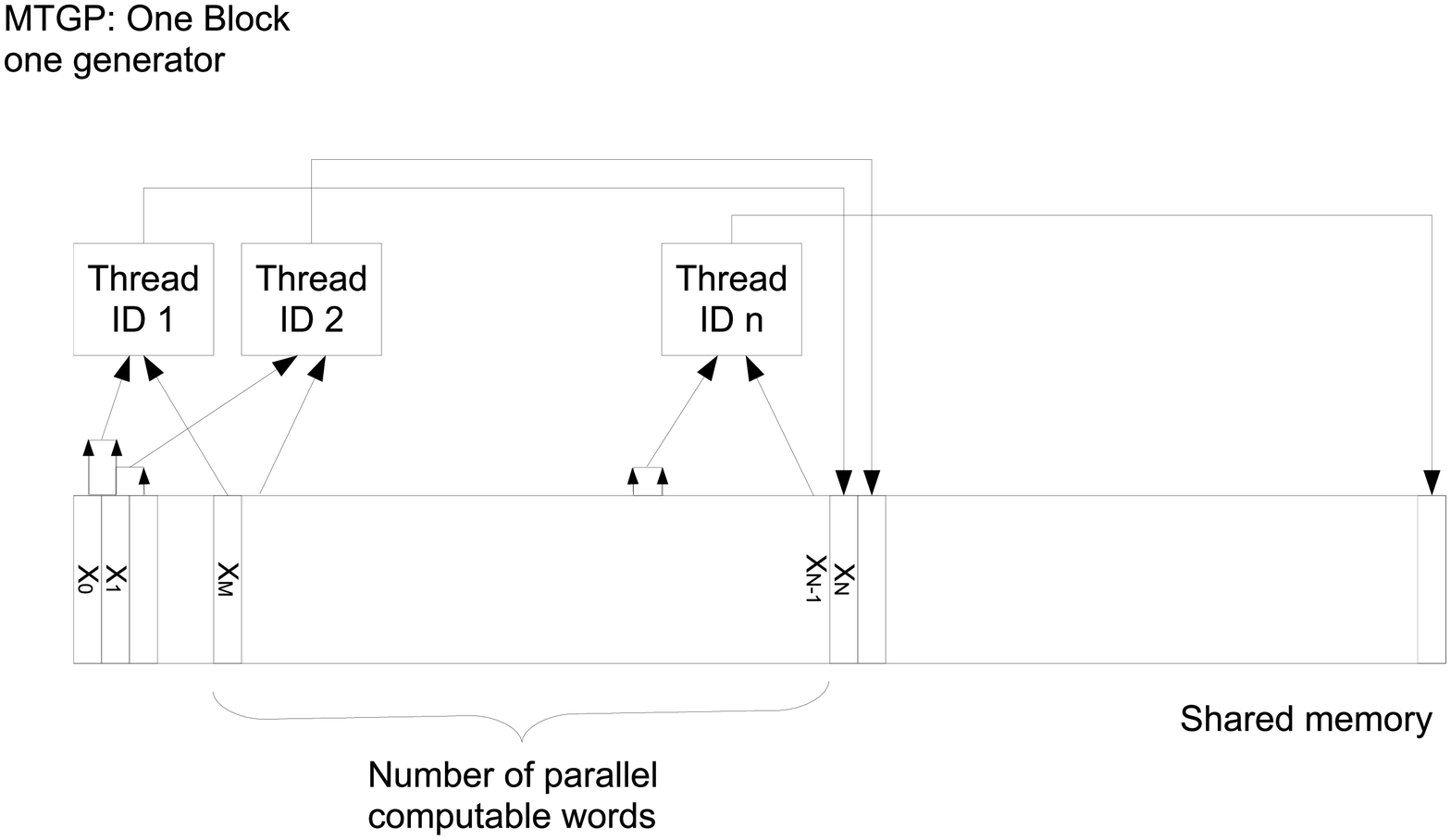}
    \caption{Strategy of MTGP: one block works on a
             single recursion using up to $N-M$ threads.
             At Step 1, the threads read from memory via
             arrows numbered 1 in parallel,
             then at Step 2 (respectively 3)
             via arrows numbered 2 (respectively 3).
             At Step 4, the threads write the generated numbers via arrows numbered 4.}
    \label{fig:block-mtgp}
  \end{center}
\end{figure}

A basic strategy in our proposed MTGP
is to use this parallelism for threads in
one block, as pictured in Figure~\ref{fig:block-mtgp}.
One block works on a single recursion using
up to $N-M$ threads in parallel.
Suppose that one block has $n$ threads
($n \leq N-M$). Then, the $i$-th thread
$(1 \leq i \leq n)$
works as follows:
\begin{description}
\item[Step 1] reads $X[i-1]$ ($1 \leq i \leq n$).
\item[Step 2] reads $X[i]$ ($1 \leq i \leq n$).
\item[Step 3] reads $X[M+i-1]$ ($1 \leq i \leq n$).
\item[Step 4] computes $f(X[M+i-1], X[i], X[i-1])$
  and writes the result to $X[N+i]$ ($1 \leq i \leq n$).
\end{description}

This idea of {\em one block for one generator}
is in contrast to a simpler
idea: {\em one thread for one generator},
adopted in CUDA SDK Mersenne Twister sample,
which will be explained in \S\ref{sec:sdk-mt}.

\subsection{$\F2$-linear generators and Mersenne Twister}
\label{sec:MT}
We briefly recall the notion of $\F2$-linear generators,
in particular Mersenne Twister (MT)~\cite{MT} generator,
since MTGP is its variant. See \cite{F2RNG-LEcuyer} for
a general theory on $\F2$-linear
generators.

In this article, we identify the set of bits $\{0,1\}$ with
the two-element field $\F2$.  A $w$-bit integer is identified with a
horizontal vector in $\F2^{w}$, and $\oplus$ denotes the sum as vectors
(i.e., bit-wise exor). Thus, the set of
word-size integers is an $\F2$-linear space, as well as
the set of states of a memory array, etc.
We mean by an $\F2$-linear generator
a pseudorandom number generator with $\F2$-linear
vector state space, $\F2$-linear transition function,
and $\F2$-linear output function.

\subsubsection*{Notation}
The following notations on bit-operations on words are
used. The bitwise-and is denoted by $\&$. Let $r$
be an integer with $1 \leq r \leq w$ and $\bx$, $\by$
be two $w$-bit words.
The $(w-r)$-bit integer consisting of the $(w-r)$
most significant bits (MSBs) of $\bx$ is denoted
by $\overline{\bx}^{w-r}$.
The $r$-bit integer consisting of the $r$
least significant bits (LSBs) of $\by$ is denoted
by $\underline{\by}^{r}$.
The $w$-bit integer obtained by concatenating
$\overline{\bx}^{w-r}$ and $\underline{\by}^{r}$
in this order is denoted by $(\bx^{w-r}|\by^r)$.
The bold $\bzero$ means the zero word.
Thus,
$(\bx^{w-r}|\bzero^r)$ is a word whose most significant $w-r$ bits
coincide with those of $\bx$ and the other $r$ bits being 0.
The logical left shift of $\bx$ by $r$ bits is denoted by
$(\bx \ll r)$, and the right shift by
$(\bx \gg r)$. A hexadecimal number is denoted with {\tt 0x}
at the head,
so for example $\tt 0xf$ denotes the integer 15 whose
binary representation is 1111.
Matrices are denoted by bold letters such as $\bA$ and $\bR$,
and multiplication to a row vector $\bx$ is denoted by
$\bx \bA$, and every component is computed modulo 2, i.e.,
in $\F2$.

Choose a Mersenne prime, i.e.,\ a prime number of
the form of $2^p-1$; the integer $p$ is called
a Mersenne exponent (MEXP).
A basic strategy of MT is to realize the Mersenne
prime period. For this purpose, put
$N = \lceil p / w \rceil$, $r = wN - p$.
Thus, $N$ is the least length of array of
$w$-bit integers that accommodates $p$ bits.
MT generates a sequence
of elements in $\F2^{w}$ by a recursion
\begin{eqnarray}
\bx_{N+i}&=&f(\bx_{M+i}, \bx_{1+i}, \overline{\bx_{i}}^{w-r}) \label{eqn:mth}\\
       &=&\bx_{M+i} \oplus (\bx_{i}^{w-r}|\bx_{1+i}^r)\bA, \label{eqn:mtmat}
\end{eqnarray}
where
$\bA$ is a $(w \times w)$-matrix such that
$\bx \bA$ is computable by
\begin{equation}\label{eq:multiplication-by-A}
\bx \bA = \left\{\begin{array}{ll}
           (\bx \gg 1) & (\mbox{if $\underline{\bx}^1=0$}) \\
           (\bx \gg 1) \oplus {\ba} &
		(\mbox{if $\underline{\bx}^1=1$}),
           \end{array}\right.
\end{equation}
where $\ba$ is a constant $w$-dimensional row vector.

The function
$$
(\bx_{N+i-1},\bx_{N+i-2},\ldots,\bx_{i+1},\overline{\bx}_i^{w-r})
 \mapsto
(\bx_{N+i},\bx_{N+i-1},\ldots,\bx_{i+2},\overline{\bx}_{i+1}^{w-r})
$$
is a fixed $\F2$-linear transformation $F$, and the recursion
is simply the iteration of $F$. Thus,
Mersenne Twister is considered as an automaton
with state transition function $F:\F2^p \to \F2^p$,
where $p=wN-r$. The period of $F$ attains the maximum
$2^p-1$ if and only if the characteristic polynomial
of $F$ is primitive --- there is an efficient algorithm
to check this when $p$ is a Mersenne exponent \cite{MT}.

\subsection{$k$-distribution and tempering}
The quality of a pseudorandom number generator
is often measured by the period, together with
its high-dimensional
equidistribution property, defined below (cf. \cite{CLT}\cite{COMBTAUS}).
\begin{definition}
    A sequence of $v$-bit integers with period $P=2^p-1$
    \[
    \bx_0, \bx_1, \ldots, \bx_{P-1}, \bx_P=\bx_0, \ldots
    \]
    is said to be {\em $k$-dimensionally equidistributed}
    if the consecutive $k$-tuples
    \[
    (\bx_i, \bx_{i+1}, \ldots, \bx_{i+k-1}), \quad i=0, \ldots, P-1
    \]
    are uniformly distributed over all possible $kv$-bit patterns
    except the all-0 pattern, which occurs once less often, i.e.,
each distinct $k$-tuple of $v$-bit words appears the same number
of times in the sequence (with one exception of $k$-tuple of zeroes).
\end{definition}

\begin{definition}
    A periodic sequence of $w$-bit integers
    is {\em $k$-dimensionally equidistributed to $v$-bit accuracy}
    if the most significant $v$-bit integer sequence is
    $k$-dimensionally equidistributed.

    The {\em dimension of equidistribution to $v$-bit accuracy}
    $k(v)$ is the maximum value of $k$ such that the sequence
    is $k$-dimensionally equidistributed to $v$-bit accuracy.
    Larger values of $k(v)$ show higher
    dimensional equidistribution for $v$-bit accuracy.

    For $P=2^p-1$, there is a bound
    $
    k(v) \leq \lfloor p / v \rfloor.
    $
    The {\em dimension defect} $d(v)$ at $v$
    is the difference
    $
    d(v):= \lfloor p / v \rfloor - k(v) \geq 0,
    $
    {\em the total dimension defect} $\Delta$
    is their sum over $v$:
     $
     \Delta := \sum_{v=1}^{w}d(v).
     $
    Smaller $\Delta$ value shows that the generator
is closer to the optimum from the view point of $k(v)$
$(v=1,\ldots,w)$.
\end{definition}
A possible alternative to $\Delta$ is {\em the maximum
dimension defect} $\max \{d(v) | v=1,2,\ldots,w\}$.
In this article we use only $\Delta$.

To obtain a better equidistribution property,
MT chooses a $(w \times w)$ matrix $\bT$
(called the tempering matrix),
and outputs
$\bx_i\bT , \bx_{i+1}\bT, \bx_{i+2}\bT, \ldots$.
The tempering matrix $\bT$ is chosen so that
each $k(v)$ (in particular for small $v$)
is close to its upper bound mentioned above;
namely, so that the dimension defect $d(v)$
and their sum $\Delta$ is small.

\subsection{Design of MTGP : using threads}
In designing MTGP,
we utilize the parallelism explained in \S\ref{sec:prng-mtgp},
namely $N-M$ threads can work in parallel on the state space.
We chose
the number of threads for degree-$N$ MTGP
generators as the largest power of 2
no more than $N-2$ (here $M\geq 2$ implies $N-M\leq N-2$),
which we denote by
$\lfloor N-2 \rfloor_2=2^{\lfloor \log_2 (N-2) \rfloor}$.
We choose a small $M$ (but $M\geq 2$), so that
the gap $N-M\geq \lfloor N-2 \rfloor_2$ holds.
We released MTGP for
three MEXPs $p=11213, 23209, 44497$.
These choices correspond to the number of threads
being 256, 512, and 1024, respectively.

\subsection{Description of MTGP:recursion}
\label{sec:mtgp-rec}
Let us fix an MEXP $p$. Similarly to MT (see \S\ref{sec:MT}),
we compute $N = \lceil p / w \rceil$ and $r = wN - p$.

MTGP generates a sequence of elements in $\F2^{w}$ by a recursion
\begin{eqnarray}
\bx_{N+i}&:=&g(\bx_{M+i}, \bx_{1+i}, \overline{\bx}_{i}^{w-r}) \label{eqn:recursion},
\end{eqnarray}
where
$M$ is an integer satisfying the condition
in the previous section.
The $\F2$-linear recurring function $g$ is determined by six integer parameters
$N,M,w,r, sh1, sh2$ and a $(4\times w)$ matrix $\bR$.
\begin{definition}\label{def:recpar}
The MTGP recursion (\ref{eqn:recursion})
is defined as follows,
using temporary variables
$\bt$ and $\bu$ of $w$-bit integer (identified with row vectors in $\F2^w$):
\begin{align}
  \bt &\leftarrow \bx_{1+i} \oplus (\bx_{i}^{w-r}|\bzero^r) \nonumber\\
  \bt &\leftarrow \bt \oplus (\bt \ll sh1) \nonumber\\
  \bu &\leftarrow \bt \oplus (\bx_{M+i} \gg sh2) \nonumber\\
  \bx_{N+i} &\leftarrow \bu \oplus (\underline{\bu}^{4}\bR),
  \label{eqn:algolrec}
\end{align}
with the computation in this order.
In the last line, $\underline{\bu}^4$ denotes
the four-dimensional row vector over $\F2$ consisting
of the four LSBs of $\bu$,
and hence the multiplication
$\underline{\bu}^4 \bR$ with $(4\times w)$-matrix $\bR$
yields a $w$-dimensional vector (see notations in \S\ref{sec:MT}).

The parameters $(N,M,w,r, sh1, sh2, \bR)$ are called the
{\em recursion parameters} of the MTGP, and chosen to
realize the period $2^{wN-r}-1$.
\end{definition}
For the speed, the multiplication of $\bR$
is implemented as a look-up table. Since $\underline{\bu}^4$
may assume only 16 values $0000, 0001, \ldots, 1111$ in
the binary form, we may precompute 16 $w$-dimensional vectors
$(0000)\bR$, $(0001)\bR$, $\ldots$, $(1111)\bR$ and store them
in an array of words, say, in \texttt{rectbl[0..15]}. Then,
$\underline{\bu}^4 \bR$ is \texttt{rectbl}[$\underline{\bu}^4$].

An equivalent description to the recursion (\ref{eqn:algolrec}) in C-language
is as follows. We assume that the content of the array \texttt{rectbl} is
precomputed as above. Note that the symbol $\wedge$ denotes
the bitwise exor in C.
\begin{align}
  \texttt{t} &= \texttt{X[1+i]} \wedge
   (\texttt{X[i]} \;\&\; \texttt{BITMASK(w,r)}) ; \nonumber\\
  \texttt{t} &= \texttt{t} \wedge (\texttt{t} \texttt{<<} \texttt{sh1}); \nonumber\\
  \texttt{u} &= \texttt{t} \wedge (\texttt{X[M+i]} \texttt{>>} \texttt{sh2}); \nonumber\\
  \texttt{X[N+i]} &= \texttt{u} \wedge \texttt{rectbl}[\texttt{u} \;\&\; \texttt{0xf}];
  \label{eqn:algolrec-C}
\end{align}
Here, \texttt{BITMASK(w,r)} is the bitmask of a $w$-bit integer with
$(w-r)$ MSBs being 1 and $r$ LSBs being 0.

The cache of the texture memory of
each block is suitable for such a look-up table,
because it is free from bank-conflicts.
Note that because the values of $\underline{\bu}^4$ in distinct
threads (even in a warp) may often coincide, the execution of
(\ref{eqn:algolrec-C}) may often cause a bank-conflict
if the array \texttt{rectbl} is kept in the shared memory.
If we keep them
in registers, the number of available registers for each thread
decreases, hence the number of concurrently executable threads decreases
since there is a limitation in the total number of registers
in one block.

\subsection{Description of MTGP : tempering}
Analogous to MT, MTGP transforms the sequence $\bx_i$
by a fixed linear transformation to obtain better $k(v)$ (called {\em tempering}):
the $(N+i)$-th output $\bo_{N+i}$
is given by
\begin{equation}
  \bo_{N+i} :=\bx_{N+i} \oplus h(\bx_{M-1+i}) \label{eqn:mtgp-tempering}
\end{equation}
for a suitably chosen linear transformation $h$
so that the output sequence
has a high value of $k(v)$ ($v=1,\ldots,w$).
\begin{definition}\label{def:tmppar}
The tempering (\ref{eqn:mtgp-tempering}) of MTGP
is defined as follows,
using a temporary variable
$\bt$ of $w$-bit integer:
\begin{align}
  \bt &\leftarrow  \bx_{M-1+i} \oplus (\bx_{M-1+i} \gg 16) \nonumber \\
  \bt &\leftarrow  \bt \oplus (\bt \gg 8) \nonumber \\
  \bo_{N+i} &\leftarrow \bx_{N+i} \oplus \underline{\bt}^4 \bT
  \label{eqn:algoltmp}
\end{align}
with the computation in this order.
Here, $\bT$ is a $(4 \times w)$-matrix over $\F2$.
The defining parameter is $\bT$,
called the {\em tempering parameter}.
The linear transformation $h(\bx_{M-1+i})$ appearing in
 (\ref{eqn:mtgp-tempering})
is defined as $\underline{\bt}^4 \bT$ appearing in (\ref{eqn:algoltmp}).
\end{definition}
This type of tempering using another word $\bx_{M-1+i}$ in the state array
came from \cite{Harase2009}. Use of the multiplication by
a $(4\times w)$-matrix was obtained through trial-and-error
to obtain a fast tempering with a large value of $k(v)$
($1\leq v \leq 32$).

An equivalent description to the tempering (\ref{eqn:algoltmp}) in C-language
is as follows:
\begin{align}
  \texttt{t} &= \texttt{X[M-1+i]} \wedge (\texttt{X[M-1+i] >> 16}); \nonumber \\
  \texttt{t} &= \texttt{t} \wedge (\texttt{t >> 8}); \nonumber \\
  \texttt{return} & \phantom{=}  (\texttt{X[N+i]} \wedge
               \texttt{tmptbl}[\texttt{t} \;\&\; \texttt{0xf}]);
  \label{eqn:algoltmp-C}
\end{align}
Similarly to the look-up table \texttt{rectbl}
in the recursion (\ref{eqn:algolrec-C}),
\texttt{tmptbl} is an array of length 16 storing
the values $(0000)\bT$, $(0001)\bT$, $\ldots$, $(1111)\bT$.

\begin{figure}
  \begin{center}
    \includegraphics[width=0.8\textwidth]{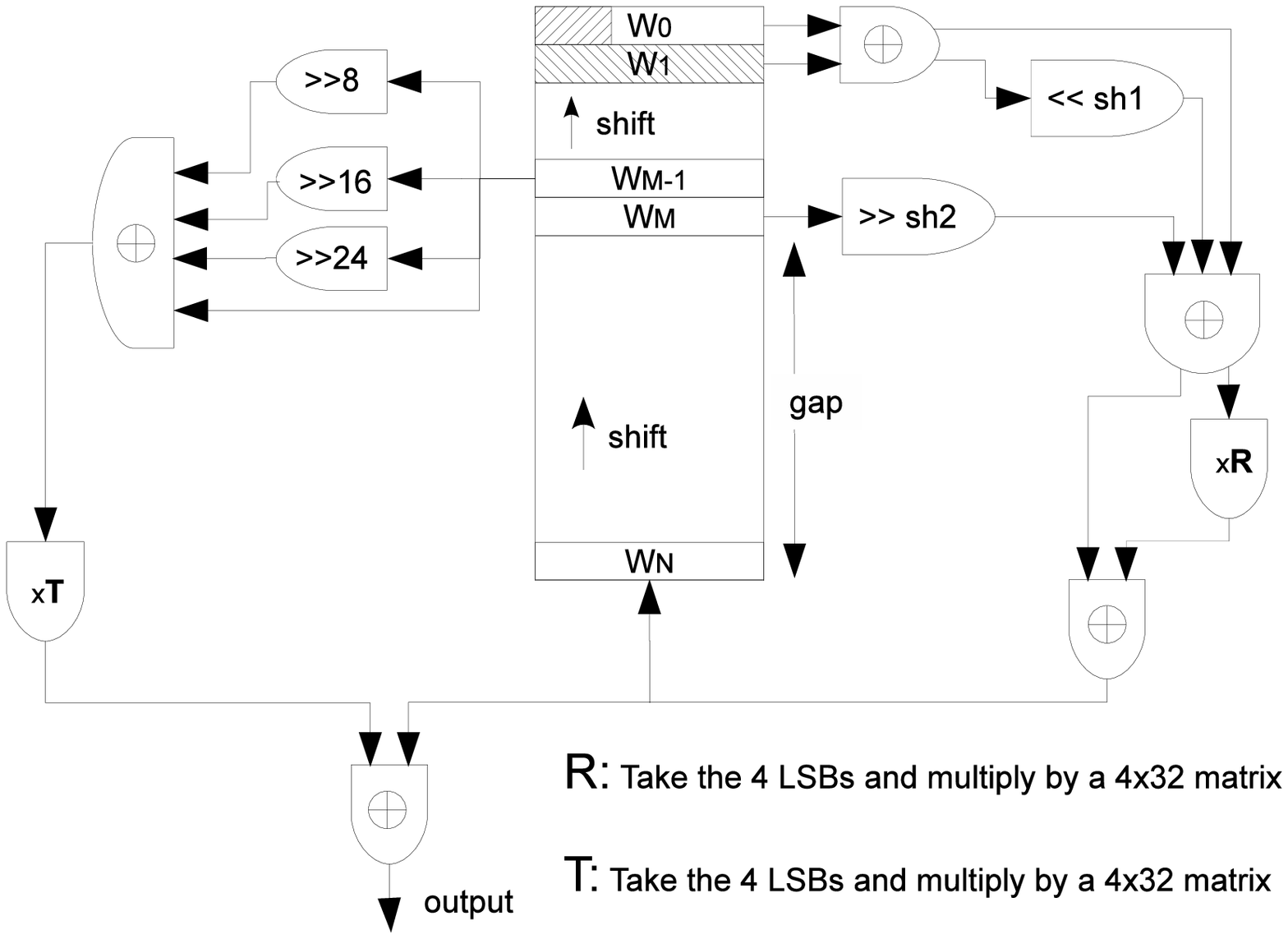}
    \caption{Circuit-like description of MTGP: the right circuit
             is for recursion, the left circuit for the tempering.}
    \label{fig:circuit}
  \end{center}
\end{figure}

Altogether, one MTGP has the recursion parameters as in Definition~\ref{def:recpar}
and the tempering parameter as in Definition~\ref{def:tmppar}.
A circuit-like description of MTGP is shown in Figure~\ref{fig:circuit}.

%
%
\subsection{Dimensions of equidistribution of MTGP}

MTGP supports three different periods $2^{11213}-1$, $2^{23209}-1$
and $2^{44497}-1$. For each period, we searched for 128
different parameter sets with good equidistribution
properties, sorted in terms of the total defect $\Delta$.
So users can generate 128 distinct streams by MTGPs of
different parameter sets.

Table~\ref{tab:de-32-23209-0} lists $k(v)$ and $d(v)$ of the MTGP23209
of period $2^{23209} -1$ for the first parameter set among
the obtained 128 sets. In addition, the defect ratio
$100\times d(v)/\lfloor p/v\rfloor$ in \% is listed.
For $v=1,\ldots,16$, $d(v)$ is 0 or 1, which
means that the equidistribution up to 16-bit accuracy
is close to the optimal.
For comparison, Mersenne Twister MT19937 has $d(1)=d(2)=d(4)=d(8)=0$,
but $d(3)=405, d(5)=249, d(6)=207$ and $d(7)=355$ for $1\leq v \leq 8$.
The total dimension defect $\Delta=1141$ of MTGP23209
is better than $\Delta=6750$ of MT19937.

The WELL \cite{WELL} generator has optimal $\Delta = 0$,
but it seems difficult
to run WELL on GPUs efficiently because of the heavy dependencies
among the partial computations in the recursion.

The last (and worst) among the 128 parameter sets
for MTGP23209 has $\Delta=2100$, still much better than MT19937.

\begin{table}
  \begin{center}
    \caption{$k(v)$, $d(v)$ and the defect ratio $r(v):=100 \times
   d(v)/(k(v)+d(v))$
(in \%) of MTGP23209
   with first parameter set}
    \label{tab:de-32-23209-0}
\begin{tabular}{|r|r|r|r||r|r|r|r|} \hline
$v$ & $k(v)$ & $d(v)$ & $r(v)$\% & $v$ & $k(v)$ & $d(v)$ & $r(v)$\% \\ \hline
1 & 23209 & 0 & 0 & 17 & 1362 & 3 & 0.22 \\ \hline
2 & 11604 & 0 & 0 & 18 & 1266 & 23 & 1.78 \\ \hline
3 & 7736 & 0 & 0 & 19 & 1181 & 40 & 3.28 \\ \hline
4 & 5802 & 0 & 0 & 20 & 1137 & 23 & 1.98 \\ \hline
5 & 4641 & 0 & 0 & 21 & 1043 & 62 & 5.61 \\ \hline
6 & 3868 & 0 & 0 & 22 & 931 & 123 & 11.67 \\ \hline
7 & 3315 & 0 & 0 & 23 & 930 & 79 & 7.83 \\ \hline
8 & 2900 & 1 & 0.03 & 24 & 930 & 37 & 3.83 \\ \hline
9 & 2578 & 0 & 0 & 25 & 727 & 201 & 21.66 \\ \hline
10 & 2320 & 0 & 0 & 26 & 726 & 166 & 18.61 \\ \hline
11 & 2109 & 0 & 0 & 27 & 725 & 134 & 15.60 \\ \hline
12 & 1934 & 0 & 0 & 28 & 725 & 103 & 12.44 \\ \hline
13 & 1785 & 0 & 0 & 29 & 725 & 75 & 9.38 \\ \hline
14 & 1657 & 0 & 0 & 30 & 725 & 48 & 6.21 \\ \hline
15 & 1547 & 0 & 0 & 31 & 725 & 23 & 3.07 \\ \hline
16 & 1450 & 0 & 0 & 32 & 725 & 0 & 0 \\ \hline
   \end{tabular}

    Total dimension defect $\Delta$ is 1141.
  \end{center}
\end{table}

%
%
\section{Comparison with other generators on GPU}
\subsection{SDK-MT: a naive approach}
\label{sec:sdk-mt}
As mentioned earlier in \S\ref{sec:prng-mtgp},
a more naive idea for PRNG for GPU is to
use one PRNG for one thread, unlike for one block as in MTGP.
An example is SDK Mersenne-Twister (SDK-MT)
in CUDA-SDK~\cite{CUDASDK} sample programs from NVIDIA, which is a
set of 4096 (small) Mersenne Twister implementations on CUDA.
SDK-MT uses 32 blocks, each block consists of 128 threads,
and each thread runs a Mersenne Twister with 607-bit state
space (MT607). Each of the $32 \times 128=4096$ threads uses
a distinct set of parameters,
produced by Dynamic Creator for MT \cite{DC}
to support the independence of those streams.

\begin{figure}
  \begin{center}
    \caption{SDK-MT: PRNGs for GPU, one thread for one generator.}
    \includegraphics[width=0.8\textwidth]{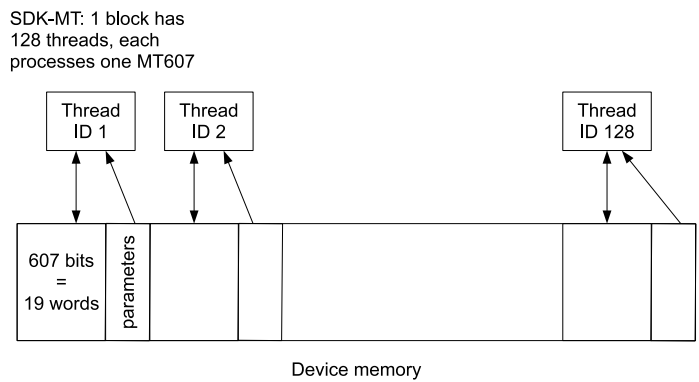}
    \label{fig:sdk-mt}
  \end{center}
\end{figure}

Figure~\ref{fig:sdk-mt}
pictures one block (128 threads) for SDK-MT.
These parameters (e.g., the coefficients in the recursion)
are kept in the global memory.

\subsection{Merits of MTGP over SDK-MT}
The parallelization in MTGP, namely
using the parallelism naturally appearing in the
recursion in the array, has been known
since the 1980s (see for example \cite[\S5.2.1]{anderson}).
Its merits compared to SDK-MT are:
\begin{itemize}
  \item SDK-MT's consumption of memory counted in bits is
    $(607+\mbox{parameter size})\times$ the number of threads.

  \item MTGP's consumption is 32$\times$ the number of threads.

  \item
     As an illustrative example, assume that the size of
     the shared memory is 16Kbyte and that
     the state spaces of SDK-MT are allocated in the shared memory.
     Then, the number of parallel threads is small: namely
     (16Kbyte)/(size of working space for MT607)$<$400, losing
     its gain in parallelism.
  \item The period of generated sequence: SDK-MT has period $2^{607}-1$,
    while MTGP has period $2^{11213}-1$ and higher dimensional
    equidistribution property.
\end{itemize}

\subsection{Other GPU-based PRNGs}
\label{sec:other-prng}
We compare MTGP with some other generators implemented for GPUs.
We do not treat classical Linear Congruential Generators with
modulus less than or equal to $2^{32}$, such as
the Park-Miller generator implemented for GPUs \cite{PARKMILLERGEN},
since they are not recommendable for massive use
(its period is $\leq 2^{32}$, and the sequence is
rejected by several simple statistical tests
called SmallCrush, see \cite{TESTU01} and \cite{WARPGEN}).
We consider the following generators.
\begin{itemize}
\item
The MTGP generator with period of $2^{11213}-1$ described above.
We use 108 blocks to run 108 MTGPs whose characteristic polynomials
are mutually distinct.


\item
The SDK-MT generator with period of $2^{607}-1$ described in
Section~\ref{sec:sdk-mt}.
\item
The HybridTaus generator \cite[Example~37-4]{HYBRIDTAUS} in GPU Gems,
with period of around $2^{121}$.
This generator outputs the exor of two 32-bit integer streams,
one generated by a combined Tausworthe generator taus88
\cite{COMBTAUS} and the other by a linear congruential generator
``Quick and Dirty.''

\item
Warp Generator \cite{WARPGEN}. This generator is based on
a sparse $\F2$-linear transition matrix $M: \F2^{32 \times 32} \to \F2^{32 \times 32}$.
Its period is $2^{1024}-1$. The space $\F2^{32 \times 32}$ is identified
with an array of 32-bit integers of length 32, so that a warp
computes the multiplication of $M$. The content of the array is
considered as 32 of 32-bit random integers.
The raw output (WarpCorrelated)
does not pass the BigCrush test suite,
but by combining with a non $\F2$-linear output
function (sum of two 32-bit integers modulo $2^{32}$), the output (WarpStandard)
passes the BigCrush test suite. In this article, the Warp Generator means WarpStandard.

\item
The CURAND generator \cite{CURAND}. This is the newest in the NVIDIA SDK.
This is an implementation of the xorwow generator, which is the combination of
an $\F2$-linear generator xorshift and a Weyl generator (i.e.,
a linear congruential generator with multiplier 1) by addition modulo $2^{32}$,
proposed in \cite{XORSHIFT-MAR}. Its period is $2^{192}-2^{32}$.
\end{itemize}

\subsection{Generation Speed}
\label{sec:speed}

\subsubsection*{Comparison of speed}
We measured the consumed time in milliseconds, for generating $5 \times 10^7$
single floating-point numbers in the range
$[0, 1)$  for MTGP11213, SDK-MT, HybridTaus,
WarpStandard, and CURAND, on the GeForce GT120 GPU (4 cores) and
the GeForce GTX260 GPU (27 cores).
For MTGP11213, we also measured the time for generating unsigned integers
and floating-point numbers in $[1,2)$
(see \S\ref{sec:floating-point} for the reason).
Table~\ref{tab:speed} shows the results.
On both GT120 and GTX260, SDK-MT is the slowest,
and WarpStandard and CURAND are faster than MTGP.
On GT120, CURAND is 2.4 times faster than MTGP, while
on GTX260 the factor decreases to 1.7.

\begin{table}
  \begin{center}
    \caption{Consumed time for $5 \times 10^7$ numbers by five
      PRNGs, their periods, and the results of BigCrush statistical
      tests: ``pass*'' means passed all the tests, ``pass''
      means passed the tests
      except those on $\F2$-linearity,
      ``reject'' means rejected systematically by at least one of
      the tests not based on $\F2$-linearity.
    }
    \label{tab:speed}
    \begin{tabular}{|l|r|r|r|r|r|r|r|} \hline
    \small
  & \multicolumn{1}{c|}{SDK-MT} & \multicolumn{3}{c}{MTGP11213}
  & \multicolumn{1}{|c|}{HybridTaus} & \multicolumn{1}{c|}{Warp}
  & \multicolumn{1}{c|}{CURAND} \\
  & float[0,1) & 32bit int & float[1,2) & float[0,1) & float[0,1)
  & float[0,1) & float[0,1) \\ \hline
  GT 120 & 50.2ms & 38.2ms & 38.9ms & 39.8ms
  & 35.4 ms& 13.8ms & 16.5ms \\ \hline
  GTX 260 & 18.6ms & 4.7ms & 4.8ms & 5.0ms & 9.3ms & 3.2ms & 3.0ms
  \\ \hline
  Period  & $2^{607}-1$ & \multicolumn{3}{c|}{$2^{11213}-1$} & $\sim 2^{121}$
          & $2^{1024}-1$ & $\sim 2^{192}$
  \\ \hline
  BigCrush & \multicolumn{1}{c|}{reject} & \multicolumn{3}{c}{pass}
  & \multicolumn{1}{|c|}{pass} & \multicolumn{1}{c|}{pass*}
  & \multicolumn{1}{c|}{reject}
  \\ \hline

    \end{tabular}
  \end{center}
\end{table}

Since CPU's and GPU's have inherently different purposes,
there may be no fair way to compare their performance,
but for a reference, we show that MTGP is faster than a PRNG on high-spec CPUs.
SFMT~\cite{SFMT}, which is one of the fastest generators on SIMD machines,
takes 25ms to generate the same $5 \times 10^7$ number of
pseudorandom 32-bit integers
on an Intel Xeon X5570 2.93GHz 4 core 2 CPU using 1 process (i.e. one core
in one CPU), while MTGP takes 4.6ms as shown in the table.
We also tried to use 4 cores of Xeon, but it turns out to be
slower than using one core. This seems due to the overhead time
for thread generation and synchronization
(we use pthread in POSIX to generate four threads in the Xeon).

We mention that there are studies on hardware implementations
of Mersenne Twisters. For example, \cite{FPGA:GPU} implemented
MT19937 on both a FPGA hardware and on a GPU. They used the 8800 GTX GPU
which is a little slower than the GTX260. Their implementation in the GPU
takes 16.9ms to generate the same number of 32-bit integers as above,
hence slower than MTGP on the GTX260.
On the other hand, their FPGA implementation is 35 times faster than their GPU implementation
and hence is much faster than MTGP on GTX260.

\subsection{Statistical tests}
We conducted TestU01 statistical test suits \cite{TESTU01}
to the MTGPs and the four other generators in the previous
section.

The Crush library in TestU01 contains 96 tests and the BigCrush library
contains
106 tests. See the user's guide for details on the tests.
The results of the tests show that MTGP, HybridTaus and Warp are not
rejected, but SDK-MT and CURAND are systematically rejected, as explained below.
\begin{description}
\item[MTGP] Among 160 tests in the BigCrush library, four tests based on
$\F2$-linearity reject MTGPs. Two LinearComp tests
(test number 80 and 81), which measure linear complexity
of the sequence, reject all the MTGPs.
Two MatrixRank tests (number 70 and 71), which measure
the $\F2$-linear rank of 5000$\times$5000 matrices
generated from the output bit stream, reject the MTGPs
with a state size less than 5000 bits.
This rejection is common among $\F2$-linear generators
such as Mersenne Twister (see \cite{TESTU01})
and WELL generators \cite{WELL}. Because these statistics
are based on $\F2$-linear relations of each bit of the outputs,
the observed deviation would hardly cause problems when
the sequence is used as real numbers or integers
in a simulation. Other tests in the BigCrush library show no systematic
deviation of MTGPs (including those generated by MTGPDC
explained in the next section).

\item[HybridTaus]
Similarly to MTGPs, HybridTaus passed the tests except
MatrixRank tests and LinearComp tests in the BigCrush.

\item[SDK-MT]
SDK-MT is rejected by RandomWalk tests
(number 76) in addition to the above mentioned $\F2$-linearity tests
in the BigCrush.
It passes the other tests.
\item[CURAND]
CURAND is systematically rejected by three tests in the BigCrush:
CollisionOver (number 7), SimpPoker (number 27) and
LinearComp (number 81). The $p$-values are outside the interval
$[10^{-15}, 1-10^{-15}]$.
This phenomenon has been reported by \cite{FABIEN}.
Since the first test is on the random numbers
in the interval $[0,1)$, the observed deviation may cause
some erroneous results in a serious simulation.

The defect of CURAND becomes more serious when we examine the
differences between the successive outputs: namely,
from the output sequence $x_n \in [0,1)$, we make
a new sequence $y_n:=x_n-x_{n-1} \mod 1 \in [0,1)$.
Among the BigCrush battery,
three CollisionOver tests
(number 7, 8, 10),
one Gap test (number 36),
one RandomWalk test (number 75),
and one Run of Bits test (number 102)
show $p$-values outside $[10^{-15}, 1-10^{-15}]$
for $y_n$.

Even among the Crush battery,
two CollisionOver tests (number 7, 8),
one ClosePairsNJumps test (number 20),
and one Gap test (number 32)
show $p$-values outside
$[10^{-3}, 1-10^{-3}]$
systematically.

\item[Warp]
WarpStandard generator passes all the tests in the BigCrush library.

\end{description}

%
%
\section{MTGPDC}
\label{algorithm-mtgpdc}
Dynamic Creation of pseudorandom number generators~\cite{DC} is
proposed for large scale parallel simulations in which a number of
PRNGs with distinct parameters are desired.
Dynamic Creator for Mersenne Twister
(MTDC) is released online:
\url{http://www.math.sci.hiroshima-u.ac.jp/~m-mat/MT/DC/dc.html}.

MTDC receives a 16-bit integer called {\em ID},
and generates a parameter set
for a Mersenne twister with specified (Mersenne prime) period,
with the ID embedded in the recursion, so that distinct ID
assures a distinct (hence relatively prime) irreducible
characteristic polynomial of the transition function of the generator.

One problem of MTDC is that there are some IDs where
MTDC can not find a parameter set with the specified
period. This phenomenon is observed only when $w=31$.
The ID is embedded as the 16 MSBs of
the parameter $\ba$ of MT (see \S\ref{sec:MT}), and MTDC searches for
the remaining bits of $\ba$ that attain the
maximal period, but sometimes no such bit-pattern
exists. Another problem is that the 16-bit space for the ID is
somewhat narrow for the recent trend of high parallelism.

Our proposed MTGPDC receives
a 32-bit integer as an ID, which is embedded in
the $(4\times 32)$ matrix parameter $\bR$ in Definition~\ref{def:recpar}.
Although there is no assurance that distinct IDs
give distinct characteristic polynomials,
there is a heuristic argument that with high probability they would do so,
since the 32 bits are mapped to $p$ (say, 11213) bits of the
coefficients of the characteristic polynomial in a complicated way.
For the sake of completeness, MTGPDC calculates the SHA1 digest
of the coefficients of the characteristic polynomial, so that
users can check whether the characteristic polynomials
are different (by checking that the SHA1 digests are different)
if they want.

As pointed out by a referee, there is no theoretical assurance for
the independence of the generated streams even if the characteristic
polynomials
are distinct. However as far as we know, there are no reports on the
deviation of the correlations among two such $\F2$-linear generators.
We also tested a merged sequence
from the outputs of two parameter sets of MTGPs of the same period using
the BigCrush library, but
no deviation is observed.

As explained in Definitions~\ref{def:recpar} and
\ref{def:tmppar}, MTGP has two kinds of parameter sets,
namely recursion parameters and tempering
parameters. Recursion parameters decide the recursion
and hence the period of the generator.
The tempering parameter $\bT$ changes the output function
and decide the dimensions $k(v)$ of equidistribution.

\begin{figure}
  \includegraphics[width=0.8\linewidth]{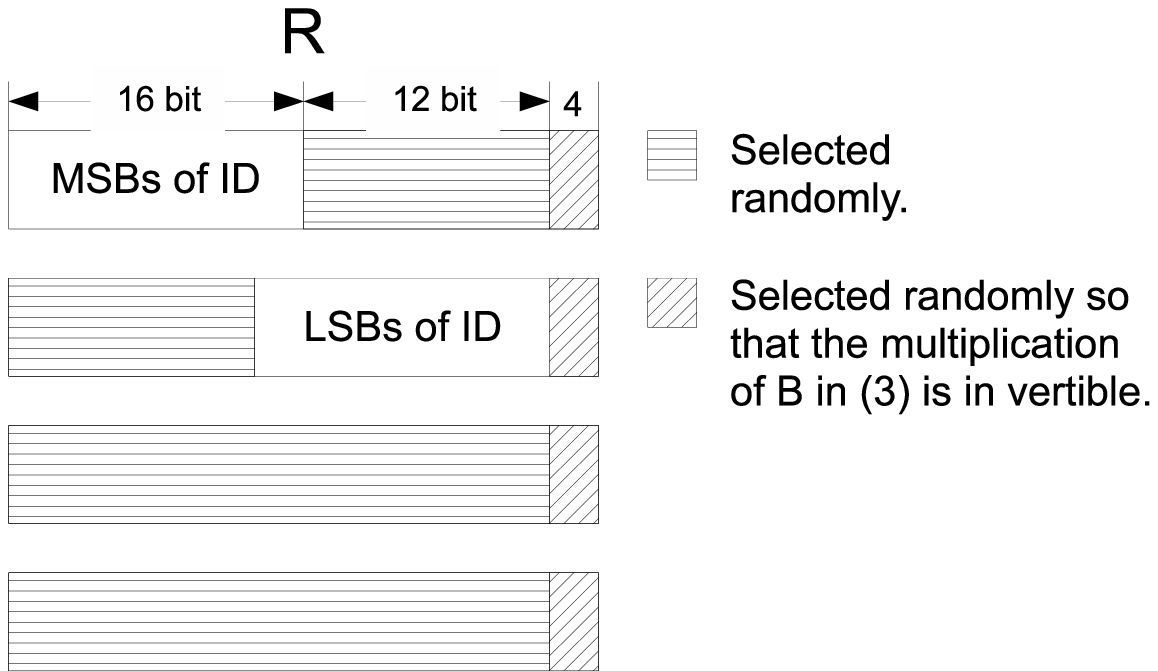}
  \caption{A recursion parameter $R$.
 For the efficiency of the search, the four sets of 4 LSBs in $R$
 are chosen so that the corresponding $4\times 4$ matrix does not have
 an eigenvalue 1, which is equivalent to the invertibility of the
 transformation on $\bu$ in the last line of (\ref{eqn:algolrec}).
 This is a necessary condition to have the maximal period.}
  \label{fig:recparam}
\end{figure}


MTGPDC first searches for the recursion parameter set
$(N,M,w,r, sh1, sh2, \bR)$ in Definition~\ref{def:recpar},
and then the tempering parameter set (Definition~\ref{def:tmppar}).
Once the Mersenne prime period $2^p-1$ and the wordsize $w$
are fixed, then $N=\lceil p / w \rceil$ and $r=Nw-p$ are determined.
MTGPDC embeds the
received ID in the matrix $\bR$, as follows.

Figure~\ref{fig:recparam} describes the recursion parameter $\bR$,
which consists of four 32-dimensional row vectors over $\F2$.
These vectors are identified with 32-bit integers,
with the MSB at the left-most component of the vector.
The given 32-bit ID is separated into the 16 MSBs
and the 16 LSBs. The former are
embedded in the 16 MSBs of the first row.
The latter are embedded in the 16 bits of the
second row consisting of the 12th, 13th, $\ldots$, and 28th MSB.
4 LSBs of each row are selected randomly so that
the corresponding $(4 \times 4)$ matrix plus the identity
matrix is invertible (this is a necessary condition
for the last line in (\ref{eqn:algolrec}) being invertible,
which is necessary for the irreducibility of
the characteristic polynomial).
Other parts of matrix $R$ are randomly selected.

Shift parameter \texttt{sh1} and \texttt{sh2} are fixed to
13 and 4 respectively, and the middle position $M$ is
randomly selected so that $2 < M < N - \lfloor N \rfloor_2$.
Experimentally, we found that
some shift parameters gave rather large defect $\Delta$ even after
tempering, so we decided to fix these
shift parameters which give good $\Delta$ empirically.
For each selection of a parameter set, the characteristic polynomial is
calculated using Berlekamp-Massey algorithm implemented
in Number Theory Library~\cite{web:NTL}. If the polynomial is
reducible, the next possible parameter set (with ID still embedded in $\bR$)
is randomly generated, until the irreducible polynomial is found.
Since the degree of the irreducible polynomial is a Mersenne exponent,
the polynomial is primitive and the maximal period $2^p-1$ is attained.
According to our experiments, probably due to the
large degree of freedom in $\bR$, MTGPDC finds a parameter set
for any given ID, unlike the case of MTDC which
fails to find one for some ID when $w=31$.

Once the recursion parameter set with maximal period is found,
MTGPDC searches for a tempering parameter set.
The tempering parameter is a $4 \times 32$ matrix (realized by
a look-up table \texttt{tmptbl}) which consists of four 32-bit vectors
(see (\ref{eqn:algoltmp})).
MTGPDC searches for tempering parameters as follows:

\begin{algorithmic}
  \STATE $\bullet$ prepare an array tmp[0..3] of 32-bit integers.
  \STATE $\bullet$ set the four components of tmp[0..3] to 0.
  \STATE $\bullet$ search for the 23 MSBs of the parameters
  \FOR{$i = 0$ to 3}
  \FOR{$j = 0$ to 20 step 5}
  \STATE $\bullet$ $e = \text{min}(j + 5, 23)$
  \STATE $\bullet$ generate all bit patterns of $j$-th to $e$-th bits of
tmp[$i$].
  \STATE $\bullet$ calculate $k(1), k(2), \ldots k(e)$ for each bit pattern.
  \STATE $\bullet$ fix the bit pattern that attains the least sum
     of the dimension defects $d(1)+d(2)+\cdots+d(e)$.
  \ENDFOR
  \ENDFOR
  \STATE $\bullet$ modify the 9 LSBs of the parameters found above
     according to the distribution of 9 LSBs
  \FOR{$i = 0$ to 3}
  \FOR{$j = 0$ to 9 step 5}
  \STATE $\bullet$ $e = \text{min}(j + 5, 9)$
  \STATE $\bullet$ generate all bit patterns of 31-$j$-th to 31-$e$-th
bits of tmp[$i$]. This does not change the 23 MSBs of the parameters.
  \STATE $\bullet$ calculate $k'(1)$ to $k'(e)$,
   where $k'(v)$ means the dimension of equidistribution for the $v$ LSBs.
  \STATE $\bullet$ fix the bit pattern that attains the least sum
     of the dimension defects $d'(1)+d'(2)+\cdots+d'(e)$,
   where $d'(v)$ means the dimension defect for the $v$ LSBs.
  \ENDFOR
  \ENDFOR
\end{algorithmic}

Computing the dimension of equidistribution is
a time-consuming part in MTGPDC.
We used the SIS algorithm \cite{SIS} to compute these dimensions,
which reduces the computation time considerably.
(Note that a faster method \cite{PIS} has recently been developed.)

\begin{figure}
 \begin{center}
  \includegraphics[width=0.7\linewidth]{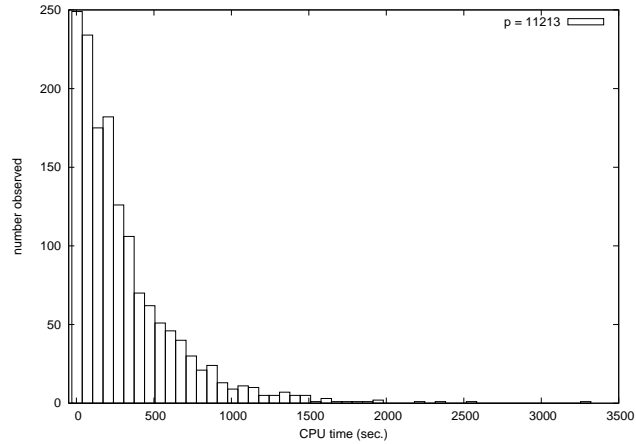}
  \caption{Distribution of time (sec) for recursion parameter search ($p = 11213$)}
  \label{fig-mtgpdc-search}
 \end{center}
\end{figure}

\begin{table}
 \begin{center}
  \caption{Time (sec) for recursion and tempering parameter search}
  \label{tab-mtgpdc-search}
  \begin{tabular}{c|lrrrrr} \hline
     & $p$ & 3217 & 4423 & 11213 & 23209 & 44497\\
     & samples & 3000 & 3000 & 1500 & 1500 & 750 \\ \hline
    recursion & min & 0 & 0 & 4 & 24 & 143 \\
     & max & 90 & 191 & 3318 & 10146 & 49987 \\
     & average & 11.2 & 25.0 & 338.1 & 1404.7 & 6529.4\\ \hline
    tempering & min & 10 & 15 & 76 & 379 & 946 \\
     & max & 25 & 40 & 253 & 1040 & 3893 \\
     & average & 21.7 & 34.1 & 213.7 & 910.0 & 3236.4\\ \hline
  \end{tabular}
 \end{center}
\end{table}

Figure~\ref{fig-mtgpdc-search} shows the distribution of the CPU time for
recursion-parameter searches for $p = 11213$.
The minimum and maximum $\Delta$
among 1500
parameter sets are 565 and 3542, respectively.
Table~\ref{tab-mtgpdc-search} shows the minimum, maximum and average time (sec.)
to find one MTGP-parameter set for various $p$.
The upper three rows show CPU times to find recursion-parameter sets and
the lower three rows show CPU times to find tempering-parameter sets.
Times in Figure~\ref{fig-mtgpdc-search} and Table~\ref{tab-mtgpdc-search}
are measured on an Intel Xeon 5500 2.26GHz 4 core $\times$ 2 CPU,
running 15 MTGPDC processes simultaneously.
These timings show that MTGPDC is too slow to use during a simulation.
MTGPDC should be used to create
a sufficient number of parameter sets before the simulation,
and these sets should be reused.

MTGPDC records a set of recursion and tempering parameters in comma-separated
values format in a file, together with the following additional data:
SHA1 digest of the characteristic polynomial, number of non-zero
terms of the characteristic polynomial, and the total dimension defect.

\subsection*{Detectable deviations in LSBs in an older version of MTGPDC}
\cite{Jonathan} reported the following problem on
the former version of MTGPDC.
In MTGPDC version 0.2 (released on June 8th 2009) for $w=32$,
several (say seven) LSBs of 32-bit outputs are often
rejected by some tests in the BigCrush library for some parameter sets.
Among these tests, two notable tests are {\tt sknuth\_Gap} test for
5-bit sequences consisting of the 26th, 27th, 28th, 29th, and 30th bit
of 32-bit outputs
(Test~35) and {\tt sstring\_HammingIndep} test for the same bits
(Test~100).
Test~35 rejects 1/5 of the 10000 created parameter sets,
while Test~100 rejects 1/10.

This older MTGPDC has only 23 MSBs for the tempering parameters.
After this report, the remaining 9 LSBs in the tempering parameters
are supplemented, as described in this section.

%
%
\section{Floating point generation in IEEE745}
\label{sec:floating-point}
If one generates random integer sequences and transforms
them into uniform real numbers by multiplication or
division, the conversion time is not negligible.
For current NVIDIA CUDA-enabled GPUs, such conversion is not heavy
(equivalent to 5 additions), but for present-day AMD GPUs,
the integer-floating conversion is rather time-consuming.

These days, most computers use IEEE 754 format
for both single and double precision floating-point numbers.
Generating floating points in this format is faster
than conversion, see \cite{DSFMT}.

The same idea is used in MTGP. IEEE 754 single-precision
format uses 32 bits for representing a real number.
It consists of
a 1-bit sign part (the MSB), a 8-bit exponent part and a 23-bit fraction part.
To generate single precision floating point numbers,
the last line of the tempering (\ref{eqn:algoltmp}) is changed to:
\begin{align}
  \bo_i &= (\bx_{i} \gg 9) \wedge
  \texttt{sngltbl}[\texttt{t} \;\&\; \texttt{0x0f}];
  \tag{C} \label{eqn:algolsingle}
\end{align}
where \texttt{sngltbl} is essentially
the same as \texttt{tmptbl}, just formatted
to produce single floating point format in IEEE754, as explained below.
Every component
\texttt{sngltbl[i]} ($0\leq i \leq 15$)
has nine MSBs which are equal to 001111111,
and its 27 LSBs are
the 27 MSBs of \texttt{tmptbl[i]}.
The above (\ref{eqn:algolsingle})
amounts to producing a 32-bit integer sequence
whose nine MSBs are 001111111 and 27 LSBs are
the 27 MSBs of the 32-bit integer version of
MTGP. The sign-bit and exponent-bit
imply that the represented real number
is in the range [1, 2), and the 27 uniformly random LSBs
imply the uniformity in that range.
By subtracting 1.0, we obtain uniform distribution in [0, 1).
It is often the case that we can use [1, 2)
directly, for example
the Box-Muller transformation for Gaussian distribution
(see~\cite[\S2]{DSFMT}).

\section{Concatenating outputs of several generators}
Occasionally practioners ask us whether it is possible to
generate a stream of a single generator (say, of MT19937)
using many processors. It would be possible by using
jumping-ahead. Here, ``jumping-ahead by $N$ steps'' means
to compute the state after
generating $N$ outputs from the present state.
If the state has $p$ bits,
then jumping-ahead costs $O(p^2)$ operations.
Even after some improvements, one jumping-ahead
costs a few milliseconds for $p=19937$
(\cite{HARAMOTO-INFORMS}), which seems too costive.

They worry that concatenating outputs of several
sequence generators may result in a poorly random sequence,
since the assurance of
the dimension of equidistribution $k(v)$ will be lost.

In our experience, concatenating the output sequences
of good pseudorandom number generators does not cause trouble.
We think the dimension of equidistribution $k(v)$
is often mis-interpreted. The $k(v)$ is the dimension
of uniformity for the whole period, and it gives no
assurance for subsequences. However, it gives assurance
on nonexistence of a linear relation:
among the consecutive $k$-tuples of $v$ bits,
there is no $\F2$-linear relation which holds forever.
This property is inherited by the subsequences, and is preserved
when concatenating outputs of several generators with
the same property.
In addition, in the cases of DC or MTGPDC, the
generators have distinct irreducible characteristic
polynomials. This property does not necessarily assure the absence of
correlations among these generators, but the risk of correlations
would be smaller than the case where generators share
one same recursion and have different seeds.

\section{Implementation using CUDA}\label{app:C}
To use (the present version of) MTGP in a GPGPU program, one needs to write
the code in CUDA~\cite{CUDA2}.

\begin{figure}
  \includegraphics[width=1.0\linewidth]{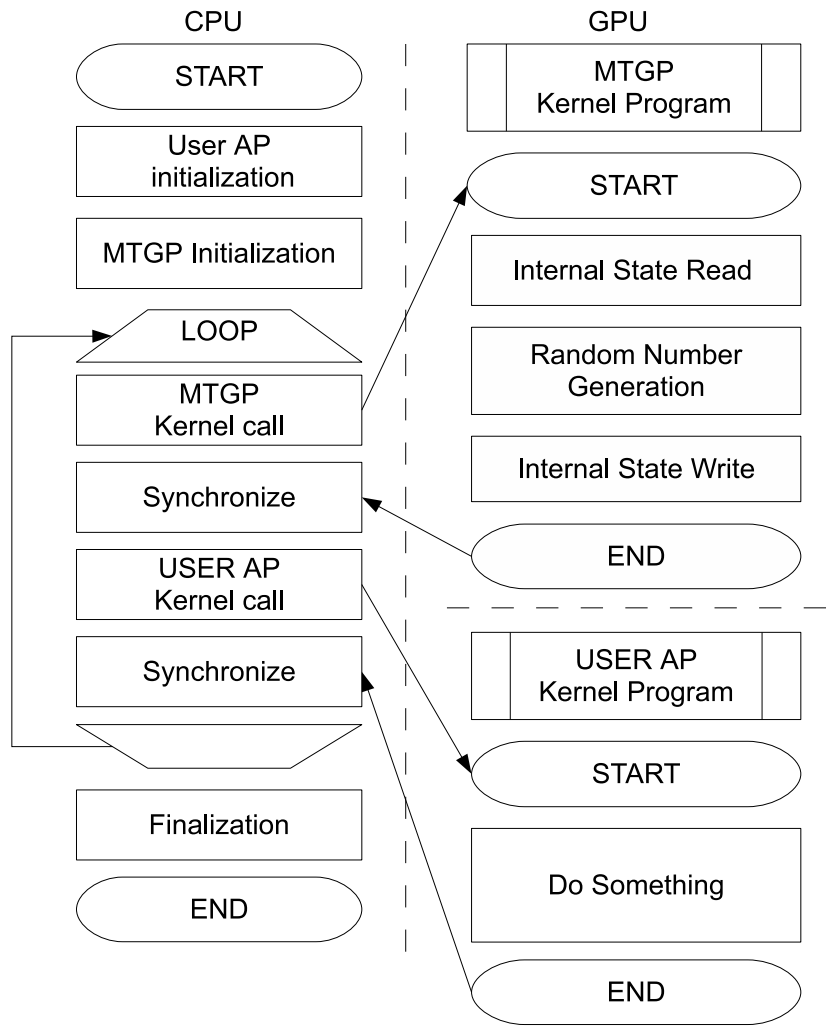}
  \caption{Flow chart for MTGP host, kernel programs and application program.}
  \label{fig-flow}
\end{figure}

A user needs to write both a host program (processed in the CPU)
and a kernel program (processed in the GPU). A flow chart is in Figure~\ref{fig-flow}.
The left hand side is the host program. User application program
(User AP) means the code which the user wants to run on GPGPU,
using the pseudorandom numbers generated by MTGP.
In the host program, initialization (or set up) of the GPU for the User
AP kernel program is executed. Then, the initialization of the GPU
for MTGP (such as texture memory set up for look-up tables,
initialization of the state array in the global memory) is executed.
Then, the host program calls the MTGP-kernel program (processed in
the GPU). In the kernel call, the host specifies (1) number of blocks
and (2) number of threads in a block, and the following arguments are passed
to the kernel program (or equivalently to each block)
(3) the number of pseudorandom numbers to be generated for each thread (4) a
pointer to global memory pointing to the state array (5) a pointer to global
memory pointing to the area where the generated numbers are stored.
When the GPU is called, each block reads the state
from global memory to shared memory, then each thread generates pseudorandom numbers
and writes them in the global memory. After writing
a specified number of outputs, each block stores the present state back
into the global memory, so that in the next MTGP kernel call
the generation continues. To have a consecutive array of pseudorandom
numbers, we need to wait for termination of all the blocks. This is done
in the host program, using the {\em synchronization} function.
Then, the User AP kernel program is called from the host.
The User AP kernel program is executed on the GPU, using
the array of generated pseudorandom numbers generated by MTGP.

One drawback of this scheme is that the user needs to
specify the number of pseudorandom numbers to be generated,
before calling the User AP kernel.

%
%
\section{Conclusions}
\label{sec:conclusion}
We proposed MTGP pseudorandom number generators
suitable for graphic processing units.
The strategy to raise its efficiency is to apply many parallel threads
to a single state space of a large pseudorandom
number generator, to realize a large state space (and hence
a long period and high-dimensional equidistribution) without
loss of its generation speed. The state space is allocated
in the shared memory in the GPU, and its parallelism of
memory access is suitable for the design of the banks
of the memory in the GPU. The generation speed of MTGP is comparable to
some of the fastest generators for the GPU. For example,
CURAND generator is faster than MTGP
by a factor of 1.7, but statistical tests show some weaknesses of CURAND.
On the other hand, WarpStandard is also faster than MTGP
and passes the BigCrush statistical test suites.
>From the viewpoint of the period and high-dimensional equidistribution,
MTGPs have better assurance than other generators.

We also designed a parameter generator for MTGP, named MTGPDC.
It runs on CPUs, receives the period and 32-bit ID,
searches for a recursion parameter set with the ID embedded,
and then searches for a tempering parameter set to attain high-dimensional
equidistribution. This facility fits a GPGPU with
many nodes. The source code of MTGP and MTGPDC, together with
their 64-bit variants, are available from the url
\url{http://www.math.sci.hiroshima-u.ac.jp/~m-mat/MT/MTGP/}.
MTGPs passed the statistical tests in the BigCrush library, except for
those tests which measure $\F2$-linear dependency of
the output sequence. The failure in such tests is common to
the Mersenne Twister and the WELL, which would not cause a problem
in a usual stochastic simulation.

\section*{Acknowledgments}
This study is partially
supported by JSPS/MEXT Grant-in-Aid for Scientific Research
No.21654004, No.19204002, No. 23244002, No.21654017, and JSPS Core-to-Core Program
No.18005. The authors are grateful for the anonymous referees for
their invaluable comments.


\bibliographystyle{plain}
\bibliography{sfmt-kanren}

\begin{thebibliography}{10}

\bibitem{anderson}
S.~Anderson.
\newblock Random number generators on vector supercomputers and other advanced
  architectures.
\newblock {\em SIAM Review}, 32(2):221--251, June 1990.

\bibitem{CLT}
R.~Couture, P.~L'Ecuyer, and S.~Tezuka.
\newblock On the distribution of k-dimensional vectors for simple and combined
  {Tausworthe} sequences.
\newblock {\em Math. Comp.}, 60(202):749--761, 1993.

\bibitem{FABIEN}
Fabien.
\newblock {XORWOW} {L'Ecuyer} {TestU01} results, 2011.
\newblock
  http://chasethedevil.blogspot.com/2011/01/xorwow-lecuyer-testu01-results.htm%
l.

\bibitem{HARAMOTO-INFORMS}
H.~Haramoto, M.~Matsumoto, T.~Nishimura, F.~Panneton, and P.~L'Ecuyer.
\newblock Efficient jump ahead for $\mathbb{F}_2$-linear random number
  generators.
\newblock {\em Informs Journal on Computing}, 20(3):385--390, 2008.
\newblock \url{doi:10.1287/ijoc.1070.0251}.

\bibitem{Harase2009}
S.~Harase.
\newblock Maximally equidistributed pseudorandom number generators via linear
  output transformations.
\newblock {\em Math. Comput. Simul.}, 79(5):1512--1519, 2009.

\bibitem{PIS}
S.~Harase.
\newblock An efficient lattice reduction method for {${\mathbf F}_2$-linear}
  pseudorandom number generators using {Mulders} and {Storjohann} algorithm.
\newblock {\em Journal of Computational and Applied Mathematics},
  236(2):141--149, August 2011.

\bibitem{SIS}
S.~Harase, M.~Matsumoto, and M.~Saito.
\newblock Fast lattice reduction for $\mathbb{F}_2$-linear pseudorandom number
  generators.
\newblock {\em Mathematics of Computation}, 80:395--407, January 2011.

\bibitem{OPENCL}
{Khronos Group}.
\newblock {OpenCL}, 2012.
\newblock \url{http://www.khronos.org/opencl/}.

\bibitem{knuth:bible}
D.~E. Knuth.
\newblock {\em The Art of Computer Programming. Vol.2. Seminumerical
  Algorithms}.
\newblock Addison-Wesley, Reading, Mass., 3rd edition, 1997.

\bibitem{HYBRIDTAUS}
{L. Howes and D. Thomas}.
\newblock Efficient random number generation and application using {CUDA},
  2009.
\newblock \url{http://http.developer.nvidia.com/GPUGems3/gpugems3_ch37.html}.

\bibitem{PARKMILLERGEN}
W.~B. Langdon.
\newblock A fast high quality pseudo random number generator for {nVidia}
  {CUDA}.
\newblock In {\em GECCO '09 Proceedings of the 11th Annual Conference Companion
  on Genetic and Evolutionary Computation Conference}, New York, 2009. ACM.
\newblock \url{doi:10.1145/1570256.1570353}.

\bibitem{COMBTAUS}
P.~L'Ecuyer.
\newblock Maximally equidistributed combined tausworthe generators.
\newblock {\em Math. Comp.}, 65(213):203--213, 1996.

\bibitem{F2RNG-LEcuyer}
P.~L'Ecuyer and F.~Panneton.
\newblock $\mathbb{F}_2$-linear random number generators.
\newblock In {\em Advancing the Frontiers of Simulation: A Festschrift in Honor
  of George Samuel Fishman}, pages 175--200, Heidelberg London New York, 2009.
  Springer-Verlag.
\newblock C. Alexopoulos, D. Goldsman, and J. R. Wilson Eds.

\bibitem{TESTU01}
P.~L'Ecuyer and R.~Simard.
\newblock {TestU01}: {A} {C} library for empirical testing of random number
  generators.
\newblock {\em ACM Transactions on Mathematical Software}, 33(4):Article 22,
  2007.

\bibitem{GPGPU}
D.~Luebke, M.~Harris, N.~Govindaraju, A.~Lefohn, M.~Houston, J.~Owens,
  M.~Segal, M.~Papakipos, and I.~Buck.
\newblock {GPGPU}: general-purpose computation on graphics hardware.
\newblock In {\em Proceedings of the 2006 ACM/IEEE conference on
  Supercomputing, November 11-17}, New York, 2008. ACM.
\newblock \url{doi:10.1145/1188455.1188672}.

\bibitem{XORSHIFT-MAR}
G.~Marsaglia.
\newblock Xorshift {RNGs}.
\newblock {\em Journal of Statistical Software}, 8(14):1--6, 2003.

\bibitem{MT}
M.~Matsumoto and T.~Nishimura.
\newblock Mersenne twister: A 623-dimensionally equidistributed uniform
  pseudorandom number generator.
\newblock {\em ACM Trans. on Modeling and Computer Simulation}, 8(1):3--30,
  January 1998.
\newblock \url{http://www.math.sci.hiroshima-u.ac.jp/~m-mat/MT/emt.html}.

\bibitem{DC}
M.~Matsumoto and T.~Nishimura.
\newblock Dynamic creation of pseudorandom number generator.
\newblock In {\em Monte Carlo and Quasi-Monte Carlo Methods 1998}, pages
  56--69. Springer-Verlag, 2000.
\newblock \url{http://www.math.sci.hiroshima-u.ac.jp/~m-mat/MT/DC/dc.html}.

\bibitem{CURAND}
{NVIDIA}.
\newblock {CURAND} library, 2010.
\newblock
  \url{http://developer.download.nvidia.com/compute/cuda/3_2/toolkit/docs/CURA%
ND_Library.pdf}.

\bibitem{CUDA1}
{NVIDIA corp.}
\newblock {\em {NVIDIA CUDA Compute Unified Device Architecture} Programming
  Guide. Ver 1.0}.
\newblock NVIDIA, 2701 San Tomas Expressway, Santa Clara, CA 95050, 1.0
  edition, 2007.

\bibitem{CUDASDK}
{NVIDIA corp.}
\newblock {CUDA SDK} code samples, 2009.
\newblock \url{http://developer.nvidia.com/object/cuda_sdk_samples.html}.

\bibitem{CUDA2}
{NVIDIA corp.}
\newblock {\em {NVIDIA CUDA} Programming Guide. Ver 2.3.1}.
\newblock NVIDIA, 2701 San Tomas Expressway, Santa Clara, CA 95050, 2.3.1
  edition, 2009.

\bibitem{WELL}
F.~Panneton, P.~L'Ecuyer, and M.~Matsumoto.
\newblock Improved long-period generators based on linear reccurences modulo 2.
\newblock {\em ACM Transactions on Mathematical Software}, 32(1):1--16, 2006.

\bibitem{Jonathan}
J.~Passerat-Palmbach, C.~Mazel, A.~Mahul, and D.~Hill.
\newblock Reliable initialization of {GPU}-enabled parallel stochastic
  simulations using mersenne twister for graphics processors.
\newblock In {\em ESM 2010, Europ. Simul. Conf. Hasselt Univ. Belgique}, page 6
  pages, 2010.
\newblock \url{doi:10.1145/1570256.1570353}.

\bibitem{SFMT}
M.~Saito and M.~Matsumoto.
\newblock {SIMD}-oriented fast {Mersenne} twister : a 128-bit pseudorandom
  number generator.
\newblock In {\em Monte Carlo and Quasi-Monte Carlo Methods 2006}, pages
  607--622. Springer, 2008.

\bibitem{DSFMT}
M.~Saito and M.~Matsumoto.
\newblock A prng specialized in double precision floating point number using an
  affine transition.
\newblock In {\em Monte Carlo and Quasi-Monte Carlo Methods 2008}, pages
  589--602. Springer-Verlag, Dec. 2009.

\bibitem{web:NTL}
Victor Shoup.
\newblock {NTL: A Library for doing Number Theory}, 1990.
\newblock \url{http://www.shoup.net/ntl/}.

\bibitem{WARPGEN}
D.~Thomas.
\newblock Uniform random number generators for {GPUs}, 2010.
\newblock \url{http://www.doc.ic.ac.uk/~dt10/research/rngs-gpu-uniform.html}.

\bibitem{FPGA:GPU}
X.~Tian and K.~Benkrid.
\newblock Mersenne twister random number generation on {FPGA}, {CPU} and {GPU}.
\newblock In {\em Proceedings of the Adaptive Hardware and Systems Conference
  2009}, pages 460--464, Los Alamitos, CA, July 2009. IEEE.
\newblock \url{doi:10.1109/AHS.2009.11}.

\end{thebibliography}

\end{document}